\documentclass[aps,pre,reprint,superscriptaddress]{revtex4-2}
\usepackage{graphicx,color}
\usepackage{mathtools}
\usepackage[T1]{fontenc}
\usepackage{bm}
\usepackage{ascmac}
\begin{document}

\title{
Ion adsorption and zeta potential of hydrophobic interfaces
}
\author{Yuki Uematsu}\email{uematsu@phys.kyutech.ac.jp}
\address{Department of Physics and Information Technology, Kyushu Institute of Technology, Iizuka, Fukuoka 820-8502, Japan}
\address{PRESTO, Japan Science and Technology Agency, 4-1-8 Honcho, Kawaguchi, Saitama 332-0012, Japan}
\date{\today}
\begin{abstract}
Hydrophobic interfaces have unique physicochemical properties and are used in various chemical products such as food, cosmetics, soap, and medicine and technologies such as pan coating and ski wax.
In this chapter, we describe the fundamental concept of hydrophobic interfaces and explain their ion adsorption and zeta potential by using experimental data from the literature.
Thus far, these electrical properties are considered universal for solid/water, liquid/water, and gas/water interfaces;
however, a careful comparison in this chapter will reveal significant differences among them.
To confirm that the affinity of $\mathrm{H^+}$ ions for all hydrophobic interfaces is stronger than that of $\mathrm{OH^-}$ ions, more experimental data on hydrophobic liquid/water and solid/water interfaces are required.

\end{abstract}

\maketitle


\vspace{3mm}

\begin{itembox}[l]{Key points/objectives box}
	\begin{itemize}
		\item The concept of hydrophobic interfaces is broad for solid/water, liquid/water, and gas/water interfaces.
		\item A careful comparison reveals that the ion adsorption and zeta potential of hydrophobic interfaces are not universal for solid/water, liquid/water, and gas/water interfaces.
		\item More experimental data on hydrophobic liquid/water and solid/water interfaces are required to confirm whether the affinities of $\mathrm{H^+}$ and $\mathrm{OH^-}$ ions are universal for all hydrophobic interfaces.
	\end{itemize}
\end{itembox}

\section{Introduction}
Hydrophobic interfaces are ubiquitous in nature and are used in technological applications such as pan coating and ski wax \cite{IsraelachviliBook,Uematsu2019}. 
The hydrophobic interfaces generally have common properties, for example, chemical inertness, water repellency, and low friction. 
The concept of hydrophobic interfaces can be applied not only to solid/water interfaces but also to liquid/water and gas/water interfaces.
However, hydrophobic and hydrophilic fluid interfaces cannot be distinguished easily because the contact angle in these interfaces is not well defined. 
Hydrophobic interfaces have the universal property of chemical inertness against water, and thus, the surface charge is governed by the physical ion adsorption. 
When acids and bases are added to these hydrophobic interfaces, the resultant difference in the surface tension demonstrates that the physical adsorption of $\mathrm{H^+}$ ions is stronger than $\mathrm{OH^-}$ ions. 
However, the zeta potentials of polytetrafluoroethylene (PTFE) surfaces, oil droplets, and gas bubbles suggest that all hydrophobic interfaces are negatively charged.
The physical chemistry community is well aware of this contradiction of hydrophobic surfaces being electrically positively vs negatively charged, and therefore, this topic is highly debated \cite{Uematsu2020,Hassanali2020}.
Notably, the $\mathrm{H^+}>\mathrm{OH^-}$ ion affinity strength for hydrophobic interfaces has been demonstrated only for air/water interfaces thus far and not for liquid/water or solid/water interfaces.

In this chapter, we describe hydrophobic surfaces and their characteristics, ion adsorption, and zeta potentials based on experimental data from the literature.  
In particular, we discuss the differences between the solid/water, liquid/water, and gas/water hydrophobic interfaces as this aspect has not been sufficiently explored by researchers thus far.
A careful comparison of the hydrophobic interfaces reported in the literature revealed a lack of experimental data for liquid/water and solid/water interfaces, even though it is a simple and important system.

\section{Hydrophobic interfaces}

\begin{figure}
\begin{center}
\includegraphics[width = 5cm]{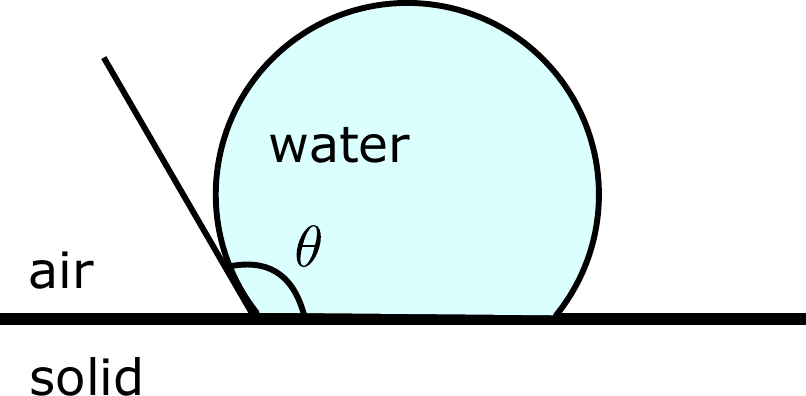}
\end{center}
\caption{Schematic of the contact angle ($\theta$) of a solid/water hydrophobic interface.
}
\label{fig:1}
\end{figure}

A flat solid surface is defined as a hydrophobic interface/surface only if its contact angle with a water droplet is greater than $90^\circ$.
The equilibrium contact angle $\theta$ of a water droplet on a flat solid surface, as shown in Fig.~\ref{fig:1}, is defined by Young’s equation, as follows:
\begin{equation}
\cos\theta = \frac{\gamma_\mathrm{s}-\gamma_\mathrm{s/w}}{\gamma_\mathrm{w}},
\end{equation}
where $\gamma_\mathrm{s}$ is the surface tension of the solid, $\gamma_\mathrm{s/w}$ is the interfacial tension at the solid/water interface, and $\gamma_\mathrm{w}$ is the surface tension of water. 
The subscript $_\mathrm{s}$ denotes a solid, and the subscript $_\mathrm{w}$ denotes water. 
Technically, $\gamma_\mathrm{s}$ and $\gamma_\mathrm{w}$ are the interfacial tensions at the solid/air and water/air interfaces, respectively, where the air phase is saturated with the vapor of the droplet solution.
However, to simplify the parameters, it is assumed that the air phase does not affect the interfacial tension, and hence, $\gamma_\mathrm{s}$ and $\gamma_\mathrm{w}$ are denoted as the surface tensions of the solid and water, respectively.  
The water surface tension is $\gamma_\mathrm{w}=72.0\,$mN/m at $1\,$atm and $298\,$K \cite{CRCBook}.
The hydrophobic condition of $\theta>90^\circ$ implies that $\gamma_\mathrm{s/w}>\gamma_\mathrm{s}$, which indicates that the interfacial tension at the solid/water interface is larger than the surface tension of the solid.

The contact angle is useful for characterizing the wetting properties of solid surfaces but is insufficient for determining both $\gamma_\mathrm{s}$ and $\gamma_\mathrm{s/w}$ separately. 
Currently, a direct method to measure $\gamma_\mathrm{s}$ and $\gamma_\mathrm{s/w}$ separately is not available, because the solid interface is not easily deformable. 
To overcome this issue, the phenomenological decomposition of the interfacial tension to the respective surface tensions is often considered \cite{VanOss1988}.
However, the decomposition of the interfacial tension is not quantitatively reliable or predictive.
Another difficulty is the contact angle hysteresis.
In general, the receding (minimum) and advancing (maximum) contact angles measured differ from the equilibrium contact angle.
The advancing contact angles of the solid surfaces are listed in Table \ref{tab:1}.
The contact angles of inert polymers exceed $90^\circ$, that of graphite is nearly $90^\circ$, and that of polymethylmethacrylate (acrylic resin) is smaller than $90^\circ$. 
Glass exhibits zero or an extremely small contact angle.

\begin{table}
\caption{Advancing contact angles and surface tension differences ($\gamma_\mathrm{s/w} - \gamma_\mathrm{s}$) of various solid materials \cite{AdamsonBook}.
$\gamma_\mathrm{s/w}-\gamma_\mathrm{s}$ is calculated using $\gamma_\mathrm{w} = 72.0\,$mN/m. }
\label{tab:1}
\begin{tabular}{ccc}
material & contact angle ($^\circ$)& $\gamma_\mathrm{s/w}-\gamma_\mathrm{s}$(mN/m)\\\hline 
paraffin & 110  & $24.6$ \\
	PTFE\footnote{polytetrafluoroethylene} & 106\footnote{average of numbers listed in pp 365 of Ref.~\citenum{AdamsonBook}.\label{average}}  & $19.8$ \\
polyethylene & 95\footref{average} & $6.3$ \\
graphite & 86 & $-5.0$ \\
PMMA\footnote{polymethylmethacrylate} & 59 & $-37.1$ \\
glass & small  & -\\
\end{tabular}
\end{table}

The definition of a hydrophobic surface based on the contact angle is applicable only for solids.
However, it can be extended to liquid/water interfaces even though contact angles cannot be measured for those.
A liquid/water interface is defined as hydrophobic only if the interfacial tension of the liquid/water interface ($\gamma_\mathrm{l/w}$) is larger than the liquid surface tension ($\gamma_\mathrm{l}$). 
For immiscible liquid/water interfaces, all the surface and interfacial tensions, $\gamma_\mathrm{l/w}$, $\gamma_\mathrm{l}$, and $\gamma_\mathrm{w}$, can be measured experimentally. 
For example, most liquid alkanes ($\mathrm{C_nH_{2n+2}}$, $5\le n\le 17$) exhibit a surface tension of approximately $\gamma_\mathrm{l}=20$ to $30\,$mN/m \cite{CRCBook,Aveyard1965} and liquid/water interfacial tension of approximately $\gamma_\mathrm{l/w}=50\,$mN/m \cite{Backes1990,Aveyard1965}, indicating that $\gamma_\mathrm{l/w}>\gamma_\mathrm{l}$; thus, the liquid alkane/water interfaces are hydrophobic.
Therefore, the effective contact angles ($\theta_\mathrm{eff}$) of these liquids can be defined as 
\begin{equation}
\theta_\mathrm{eff} = \arccos\left(\frac{\gamma_\mathrm{l}-\gamma_\mathrm{l/w}}{\gamma_\mathrm{w}}\right).
\end{equation}
Table \ref{tab:2} lists the surface tensions, interfacial tensions against water, and effective contact angles of some liquid alkanes. 
Fluorocarbon and chlorocarbon/water interfaces as well as the hydrocarbon/water interfaces are hydrophobic.  
Fatty alcohols ($\mathrm{C_nH_{2n+1}OH}$) with $n>4$ are immiscible with water; however, their effective contact angles are lower than $90^\circ$, and therefore, they are not hydrophobic.

\begin{table}
\caption{Contact angles and surface/interfacial tensions of liquid alkanes at $298\,$K and $1\,$atm.
}
\label{tab:2}
\begin{tabular}{cccc}
material 		& $\theta_\mathrm{eff}$($^\circ$) & $\gamma_\mathrm{l/w}$(mN/m) & $\gamma_\mathrm{l}$(mN/m)\\\hline 
perfluorohexane 	& $130$	& 57.2 \cite{Nishikido1989}& 11.4 \cite{Nishikido1989}\\
hexane  		& $115$	& 49.7 \cite{Backes1990} & 17.9 \cite{CRCBook}\\ 
heptane  		& $115$	& 50.1 \cite{Backes1990} & 19.8 \cite{CRCBook}\\ 
decane  		& $113$	& 52.3\footref{293} \cite{Aveyard1965} & 23.4 \cite{CRCBook}\\ 
dodecane  		& $112$ & 52.5 \cite{Aveyard1965} & 25.4\footnote{293K\label{293}} \cite{Aveyard1965}\\ 
hexadecane  		& $111$	& 53.3 \cite{Aveyard1965} & 27.1 \cite{CRCBook}\\ 
carbon tetrachloride 	& $105$	& 44.3 \cite{Backes1990} & 26.2 \cite{CRCBook}\\
polydimethylsiloxane 	& $102$ & 39.5\footref{293} \cite{Mori1984}   & 24.6\footref{293} \cite{Mori1984} \\
benzene 		& $94$ 	& 33.8 \cite{Backes1990} & 28.2 \cite{CRCBook}\\
chloroform 		& $93$	& 30.8 \cite{Backes1990} & 26.6 \cite{CRCBook}\\
1-decanol		& $75$	&  8.8\footref{293} \cite{Aveyard1977} & 28.0 \cite{CRCBook}\\
1-butanol 		& $72$	&  1.8 \cite{Backes1990} & 24.1 \cite{CRCBook}\\
\end{tabular}
\end{table}

Gas/air interfaces are unstable, and their interfacial tensions cannot be measured. 
Therefore, the definition of the effective contact angle is not applicable to gas/water interfaces.
However, all non-polar gases such as $\mathrm{N_2}$, $\mathrm{O_2}$, $\mathrm{Cl_2}$, $\mathrm{CH_4}$, $\mathrm{CO_2}$ are considered hydrophobic, even though some of them ($\mathrm{Cl_2}$ and $\mathrm{CO_2}$) react with water.
Some hydrophobic gases in air reduce the surface tension of water \cite{Mori1984, Chernyshev2014}; however, we shall ignore this effect in our discussion.

\section{Ion adsorption} 
In this section, ion adsorption at hydrophobic interfaces is discussed based on experimental surface/interfacial tension differences and contact angles.
The surface/interfacial tension increment and surface excess can be expressed by the Gibbs adsorption isotherm, as follows:
\begin{equation}
d\gamma = -\sum_i \Gamma_i d\mu_i,
\end{equation} 
where the summation is for the solute species, $\Gamma_i$ is the surface excess of the $i$th solute referenced at the Gibbs dividing surface of water, and $\mu_i$ is the chemical potential of the $i$th solute. 
Assuming that the solution is an ideal solution, the chemical potential of the $i$th solute in the bulk is expressed as
\begin{equation}
\mu_i = k_\mathrm{B}T\ln (c_i\Lambda^3),
\end{equation}
where $k_\mathrm{B}T$ is the thermal energy, $c_i$ is the bulk concentration of the $i$th solute, and $\Lambda$ is the thermal de Broglie wavelength.  
Therefore, the increment in the chemical potential can be expressed as a function of the concentration, as follows:
\begin{equation}
d\mu_i = k_\mathrm{B}T\frac{dc_i}{c_i}.
\end{equation}
Assuming that the solute is a strong electrolyte and the surface excess is linear with the salt concentration, the linear relationship between the surface/interfacial tension difference ($\Delta\gamma$) and the salt concentration ($c$) can be expressed as 
\begin{equation}
\Delta\gamma = Kc,
\end{equation} 
where $K$ is the linear coefficient of the surface/interfacial tension difference.
This simplification is not perfectly accurate; however, it is useful to discuss ion adsorption in terms of the experimental surface/interfacial tension, which often exhibits a linear dependence on salt concentration.
For instance, a negative $K$ indicates that the surface excess is positive, and thus, the electrolyte is adsorbed on the surface, whereas a positive $K$ value indicates a negative surface excess, and the electrolyte is repelled from the interface.

\begin{figure}
\includegraphics{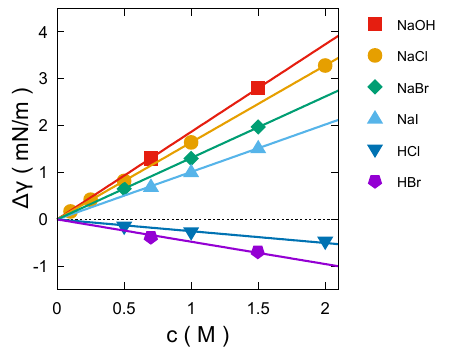}
\caption{
Experimental surface tension differences of various electrolytes.
The data points are from \cite{Washburn1930} and the linear fits have been drawn for clarity. 
	Modified with permission from \cite{Uematsu2021} Copyright 2021 IOP.
	}
\label{fig:2}
\end{figure}

Fig.~\ref{fig:2} shows the experimental data for the surface tension differences of various electrolytes \cite{Washburn1930}. 
All the electrolytes exhibit the linear dependence of surface tension differences on the salt concentrations; however, the acid electrolytes exhibit reduced surface tensions.
The separation of the individual ion surface excess is not discussed in this chapter because it is out of scope \cite{Uematsu2018,Uematsu2020}.
However, the affinity of the anions for the air/water interface is clearly in the order, $\mathrm{I^-}>\mathrm{Br^-}>\mathrm{Cl^-}>\mathrm{OH^-}$ and that of the cations is $\mathrm{H^+}>\mathrm{Na^+}$. 
By comparing the surface tensions of HCl and NaOH solutions and assuming that the magnitude of the affinities of $\mathrm{Na^+}$ and $\mathrm{Cl^-}$ for the air/water interface are the same, we can conclude that the affinity of $\mathrm{H^+}$ for the interface is stronger than that of $\mathrm{OH^-}$.
This indicates that the surface of pristine water is positively charged, which contradicts the fact that the bubbles exhibit negative zeta potentials.
The negative zeta potentials are discussed in detail in the next section.
 
\begin{figure}
\includegraphics{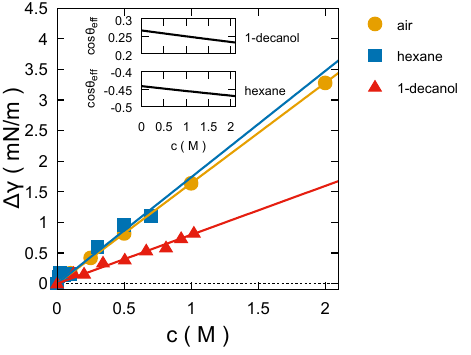}
\caption{Comparison of the surface/interfacial tensions of NaCl solution with air \cite{Washburn1930}, hexane \cite{Evans1937}, and 1-decanol \cite{Aveyard1977}.
The inset shows the effective contact angle $\theta_\mathrm{eff}$ as a function of the NaCl concentration.
}
\label{fig:5}
\end{figure}

The interfacial tension differences of liquid/water interfaces as a function of their salt concentrations have been discussed extensively in the literature \cite{Evans1937,Aveyard1976,Aveyard1977,Slavchov2014}.
Fig.~\ref{fig:5} shows the interfacial tension differences of various liquid/NaCl aqueous solution interfaces.
The hexane/water interfacial tension difference caused by the addition of NaCl is comparable to that of the air/water interface, even though the surface and interfacial tension in the absence of salts are significantly different ($\gamma_\mathrm{w}=72.0\,$mN/m and $\gamma_\mathrm{l/w} = 49.7\,$mN/m). 
The interfacial tension difference of the 1-decanol/water interface caused by the addition of NaCl is significantly lower than that of the air/water and hexane/water interfaces \cite{Slavchov2014}.
This reduction in the surface tension difference is attributed to the hydrophilic hydroxyl groups exposed to the interface \cite{Slavchov2014}.  
The inset of Fig.~\ref{fig:5} shows $\cos\theta_\mathrm{eff}$ as a function of the NaCl concentration for hexane and 1-decanol.
The variation in $\cos\theta_\mathrm{eff}$ is nearly linear with the salt concentration because the relative surface/interfacial tension differences are small.
The negative linear dependence, $d(\cos\theta_\mathrm{eff})/dc <0$, suggests that the salts make the interface more hydrophobic. 

\begin{figure}
\includegraphics{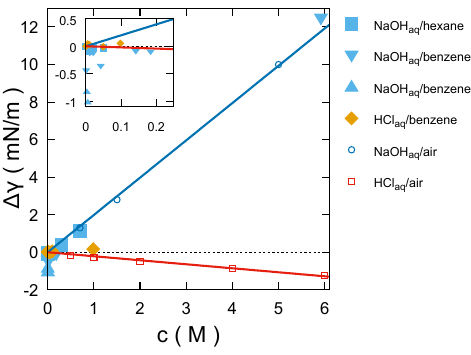}
\caption{Interfacial tension differences of liquid/water interfaces caused by the addition of acidic and basic salts. 
The data points are of the NaOH solution/hexane (filled squares \cite{Evans1937}), NaOH solution/benzene (up triangles \cite{Harkins1917}, down triangles \cite{Seetharaman2021}), and HCl solution/benzene (diamonds \cite{Harkins1917}) interfaces.
The surface tension differences of NaOH and HCl solutions (open circles and open squares \cite{Washburn1930}) and their linear fits are plotted for comparison. 
The inset is the magnification of the plot at the smaller concentrations. 
}
\label{fig:9}
\end{figure}

The affinities of cations and anions for the hydrocarbon/water interfaces are similar to those for the air/water interface \cite{Aveyard1976,Slavchov2014}.
Therefore, the order of affinity strengths is also similar \cite{Aveyard1976}.
Because the surface charge of the pristine liquid/water interface is determined by the affinities of $\mathrm{H^+}$ and $\mathrm{OH^-}$, the interfacial tension differences caused by acidic and basic salts are important.  
However, experimental data on the interfacial tension differences caused by acids and bases are limited \cite{Harkins1917,Evans1937,Seetharaman2021}.
Fig.~\ref{fig:9} shows the experimental interfacial tension differences of the liquid/water interfaces caused by the addition of acidic and basic salts.
The data of the NaOH solution/hexane interface (filled squares \cite{Evans1937}) are comparable to those of NaOH solutions (open circles \cite{Washburn1930}), whereas that of NaOH solution/benzene interface exhibited a reduced tension $\Delta\gamma<0$ (up-pointing and down-pointing triangles).
Harkins et al.~\cite{Harkins1917} reported that 5.9 M NaOH solution/benzene interface exhibited a $\Delta\gamma$ of $12.5\,$mN/m, as shown in Fig.~\ref{fig:9}, and it agrees well with the linear fit of the surface tension for NaOH solution (solid blue line).
The negative tension differences and minima at millimolar concentrations could be explained by charged impurities \cite{Uematsu2018}, and thus, the interfacial tension differences of the NaOH solution/hydrocarbon interfaces are comparable to that of the air/water interface.
The interfacial tension at concentrations not measured yet must be measured in the future to validate our explanation. 
Similarly, we speculate that the interfacial tension of the HCl solution/hydrocarbon interface is comparable to that of the air interface; however, a nearly constant interfacial tension was reported by Harkins et al. (diamonds in Fig.~\ref{fig:9}) \cite{Harkins1917}.

The interfacial tension difference of the solid/water interface cannot be measured experimentally because the interface cannot be easily deformed.
However, using the contact angle and surface tension of the electrolyte solutions, the interfacial tension difference of the solid/water interface can be calculated from the available experimental data. 
The Young’s equation for a droplet of aqueous electrolyte solution on a hydrophobic surface can be expressed as 

\begin{equation}
\gamma_\mathrm{aq}(c)\cos\theta_\mathrm{aq}(c) = \gamma_\mathrm{s}-\gamma_\mathrm{s/aq}(c), 
\label{eq:8}
\end{equation} 
where $\gamma_\mathrm{aq}(c)$ is the surface tension of the aqueous electrolyte solution, $\theta_\mathrm{aq}(c)$ is the contact angle of the droplet on the solid surface, $\gamma_\mathrm{s/aq}(c)$ is the interfacial tension of the solid/solution interface, and $c$ is the salt concentration.
For salt-free water ($c=0$), the Young’s equation is 
\begin{equation}
\gamma_\mathrm{w} \cos\theta = \gamma_\mathrm{s}-\gamma_\mathrm{s/w}.
\label{eq:9}
\end{equation} 
Subtracting eq.~\ref{eq:9} from eq.~\ref{eq:8} yields
\begin{equation}
\gamma_\mathrm{s/aq}(c)-\gamma_\mathrm{s/w}  = - \left(\gamma_\mathrm{aq}(c) \cos\theta_\mathrm{aq}(c) -\gamma_\mathrm{w} \cos\theta\right).
\label{eq:10}
\end{equation}
Therefore, by entering the values of $\gamma_\mathrm{aq}(c)$ and $\theta_\mathrm{aq}(c)$ into the equation, the interfacial tension difference of the solid/solution interface can be calculated.

\begin{figure}
\includegraphics{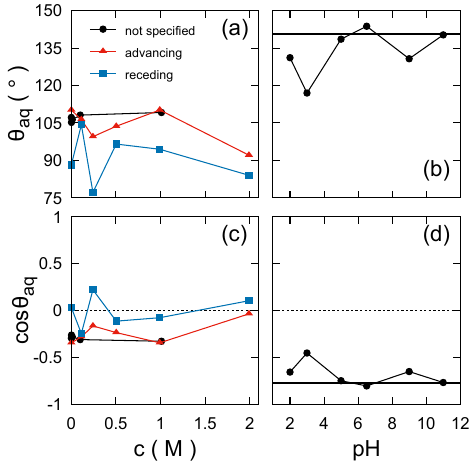}
\caption{
Contact angles of electrolyte droplets on a polytetrafluoroethylene surface \cite{Cutts2009,Gulgonul2019,Hamadi2009}.
The left panel (a,c) shows the contact angles ($\theta_\mathrm{aq}$ and $\cos\theta_\mathrm{aq}$) as a function of NaCl concentrations \cite{Cutts2009,Gulgonul2019}, whereas the right panel (b,d) shows the contact angles as a function of pH \cite{Hamadi2009}.
The solid line in (b,d) is the contact angle $\theta = 140.6^\circ$ at $\mathrm{pH}=5.6$. 
}
\label{fig:3}
\end{figure}

The contact angles of NaCl solutions on a PTFE surface are shown in Fig.~\ref{fig:3} a and c \cite{Cutts2009, Gulgonul2019}.
The contact angles reported in Ref.~\citenum{Gulgonul2019} were not categorized as advancing or receding, but they are comparable to the advancing contact angles reported in Ref.~\citenum{Cutts2009}.
Although the contact angles  are not ideal equilibrium contact angles, they remained constant for variable salt concentrations. 
Assuming that the interfacial tension of the solid/water interface is proportional to the salt concentration,  eq.~\ref{eq:10} becomes
\begin{equation}
\gamma_\mathrm{s/aq}(c)-\gamma_\mathrm{sw}  = - (K_\mathrm{aq} \cos\theta +\gamma_\mathrm{w}K_\theta) c, 
\label{eq:11}
\end{equation} 
where $K_\theta$ denotes a linear constant of the contact angle defined by
\begin{equation}
\cos\theta_\mathrm{aq}(c)\approx \cos\theta + K_\theta c.
\end{equation}
Therefore, the linear coefficient of the solid/solution interfacial tension can be obtained by   
\begin{equation}
K_\mathrm{s/aq} = -K_\mathrm{aq}\cos\theta-\gamma_\mathrm{w}K_\theta.
\label{eq:13}
\end{equation}

\begin{figure}
\includegraphics{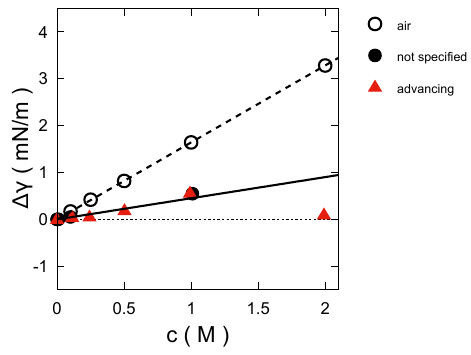}
\caption{
Calculated interfacial tension differences of PTFE/water interface caused by the addition of NaCl \cite{Cutts2009,Gulgonul2019}.
For comparison, the surface tension differences of an air/NaCl solution interface and their linear fit is plotted (open circles \cite{Washburn1930} and broken line, respectively).
}
\label{fig:4}
\end{figure}

Fig.~\ref{fig:4} shows the PTFE/water interfacial tension difference caused by the addition of NaCl, which are calculated using the contact angle data (red triangles and black circles) from Fig.~\ref{fig:3}.
For comparison, the surface tension differences of the air/NaCl solution interface are also plotted (open circles \cite{Washburn1930} and broken line). 
The solid line in Fig.~\ref{fig:4} represents eq.~\ref{eq:11} with $\theta = 107^\circ$ and $K_\theta =0$.  
Except for the value for $2\,$M NaCl, $\gamma_\mathrm{s/aq}(c)-\gamma_\mathrm{s/w}$ values increase linearly with salt concentration, which is in good agreement with the solid line obtained from the theoretical calculation (eq.~\ref{eq:11}).

\begin{figure}
\includegraphics{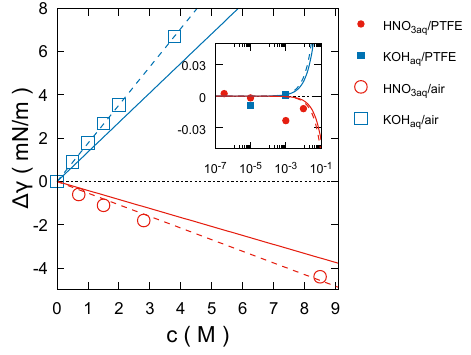}
\caption{Interfacial tension differences of solid/water interface after the addition of acidic and basic salts.
The surface tensions of $\mathrm{HNO_3}$ and $\mathrm{KOH}$ solutions \cite{Washburn1930} are denoted by open circles and open squares, whereas their linear fit is denoted by broken lines.
The solid lines are the predicted interfacial tension differences of PTFE/water interface caused by the addition of $\mathrm{HNO_3}$ and $\mathrm{KOH}$ calculated using eq.~\ref{eq:13} with $K_\theta = 0$.
The inset shows the plot of interfacial tensions calculated from the contact angles of PTFE surfaces reported by \cite{Hamadi2009} using eq.~\ref{eq:10}.
At the reference $\mathrm{pH}=5.6$ at $c=0$, the contact angle $\theta = 140.6^\circ$ is obtained by the interpolation of the data between $\mathrm{pH}=5.0$ and $6.5$.
}
\label{fig:10}
\end{figure}

The order of the affinity strengths of cations and anions for hydrophobic solid/water interfaces is important for determining the colloidal stability and surface charge density of the interface. 
However, the experimental contact angle data for various electrolytes are limited compared to the data available for the surface/interfacial tension differences of air/water and liquid/water interfaces.
In this section, we examine the interfacial tension differences of the PTFE/water interface caused by $\mathrm{HNO_3}$ and $\mathrm{KOH}$ using the contact angle data from Ref.~\citenum{Hamadi2009}. 
In Fig.~\ref{fig:3}b and d, the contact angle $\theta_\mathrm{aq}$ and its cosine ($\cos\theta_\mathrm{aq}$) of the PTFE/water interface are plotted as a function of $\mathrm{pH}$ \cite{Hamadi2009}.
The pH was adjusted to the range 2–11 by the addition of $\mathrm{HNO_3}$ or $\mathrm{KOH}$ \cite{Hamadi2009}. 
Therefore, the calculated interfacial tension differences were within a narrow concentration range.
To obtain the interfacial tension difference, the reference pH at $c=0$ was set to $\mathrm{pH}=5.6$ which is the natural pH of the air-saturated water.
The contact angle at the reference state, $\theta = 140.6^\circ$, is obtained by linear interpolation of the experimental contact angles; however, the obtained contact angle is significantly larger than those listed in Table~\ref{tab:1}. 
In Fig.~\ref{fig:10}, the surface tensions of $\mathrm{HNO_3}$ and $\mathrm{KOH}$ solutions \cite{Washburn1930} are denoted by open circles and open squares, respectively, and their linear fits are denoted by the broken lines.
The solid lines represent the predicted interfacial tension differences of the PTFE/water interface caused by the addition of $\mathrm{HNO_3}$ and $\mathrm{KOH}$ calculated using eq.~\ref{eq:13} with $\theta = 140.6^\circ$ and $K_\theta = 0$.
The inset shows a plot of the interfacial tension differences calculated using available data on the contact angles at the PTFE surface \cite{Hamadi2009}.
Because the concentration range of the reported experimental contact angles is extremely narrow, a comparison between the theoretical predicted values and the calculated experimental interfacial tension values is not conclusive. 
Contact angle measurements at higher concentrations of acidic and basic salts are necessary to clarify the affinity strength of $\mathrm{H^+}$ and $\mathrm{OH^-}$ for the PTFE/water interface. 

\section{Zeta potential}

Zeta potential is defined as the effective electrostatic potential at the interface. 
It can be determined by electrokinetic phenomena such as electro-osmotic flow, streaming current, and streaming potential measurements or 
can be calculated from the electrophoretic mobility measurements of the particles.
The zeta potentials of colloids play an important role to determine the stability of the colloidal solutions.
In this section, first, we discuss the zeta potentials of a flat solid/water interface, and then, the zeta potentials of gas/water and liquid/water interfaces derived from electrophoretic mobility.

\begin{figure}
\includegraphics[width=8cm]{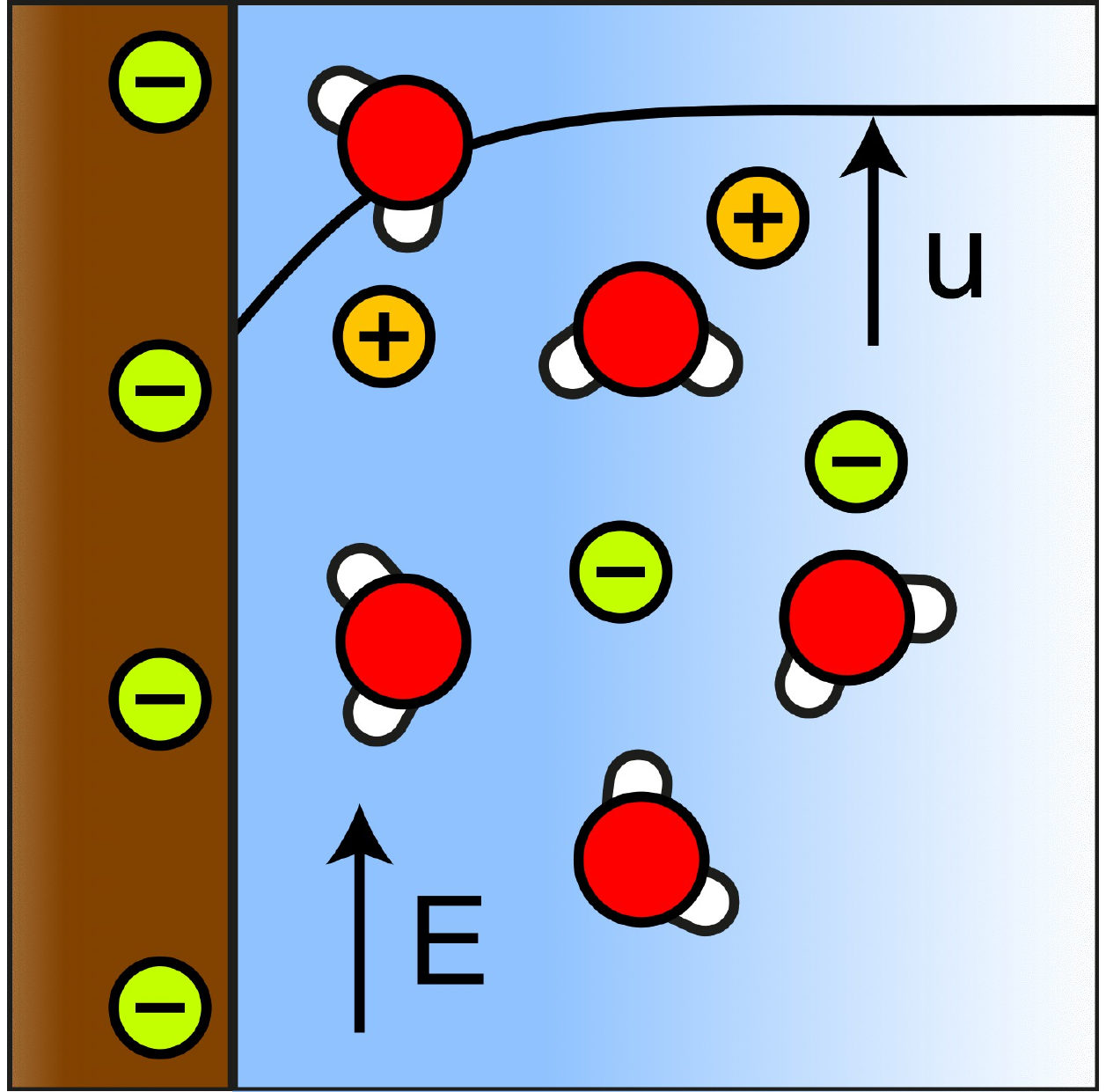}
\caption{Illustration of the electric double layer and electro-osmosis at a flat solid/water interface.
	Modified with permission from \cite{Uematsu2021} Copyright 2021 IOP.
	}
\label{fig:6}
\end{figure}

Fig.~\ref{fig:6} shows an illustration of a flat solid surface in contact with an electrolyte solution. 
The solid surface is often charged because of ion dissociation or physical ion adsorption, and counterions in the solution accumulate near the surface to form an electric double layer. 
The Poisson equation for the normal coordinate $z$ is expressed as
\begin{equation}
\varepsilon\varepsilon_0\frac{d^2\psi}{dz^2} = -\rho(z),
\label{eq:12}
\end{equation}
where $\varepsilon$ is the dielectric constant of the solution, $\varepsilon_0$ is the vacuum permittivity, $\psi$ is the local electrostatic potential, and $\rho$ is the charge density.
Electro-osmosis is defined as the flow of the solution when an external electric field ($E$) is applied along a charged surface, as shown in Fig.~\ref{fig:6}.
The velocity profile $u(z)$ of electro-osmosis can be calculated using the Stokes equation as follows:
\begin{equation}
\eta\frac{d^2 u}{dz^2} + \rho(z) E = 0,
\label{eq:13_2}
\end{equation} 
where $\eta$ is the viscosity of the solution. 
Eqs.~\ref{eq:12} and \ref{eq:13_2} are solved using the boundary conditions $\psi|_{z\to\infty}=0$ and $\psi|_{z=0}=\zeta$ to afford
\begin{equation}
u(z) = \frac{\varepsilon\varepsilon_0}{\eta}\left[\psi(z)-\zeta\right]E,
\end{equation}
where $\zeta$ is the electrostatic potential at the surface.
The electro-osmotic velocity far from the surface can be calculated using
\begin{equation}
u(z)|_{z\to\infty} = - \frac{\varepsilon \varepsilon_0 \zeta}{\eta} E,
\label{eq:15}
\end{equation}
Thus, measuring the electro-osmotic mobility $\mu = u(z)|_{z\to\infty}/E$ provides information on the electrostatic potential at the surface.
The zeta potential determines the strength of the repulsive force between charged colloids, which is important in chemical industries that manufacture food, cosmetics, soap, and medicine. 

To express the zeta potential as a function of the surface charge density, we must assume that the ions obey the Boltzmann distribution.  
Then, eq.~\ref{eq:12} can be written as the Poisson-Boltzmann equation, as follows: 
\begin{equation}
\varepsilon\varepsilon_0\frac{d^2\psi}{dz^2} = -ec\left[\mathrm{e}^{-e\psi/k_\mathrm{B}T}-\mathrm{e}^{e\psi/k_\mathrm{B}T}\right],
\label{eq:12-2}
\end{equation}
where $c$ is the bulk salt concentration, $e$ is the elementary charge, and $k_\mathrm{B}T$ is the thermal energy. 
This nonlinear differential equation is solvable; thus, the surface charge density can be expressed as
\begin{equation}
\sigma = \sqrt{8c\varepsilon\varepsilon_0k_\mathrm{B}T}\sinh\left(\frac{e\zeta}{2k_\mathrm{B}T}\right).
\end{equation}
A flat solid surface with a semi-infinite solution phase is geometrically simple and analytically tractable.
However, the measurement of electro-osmotic flow is challenging; therefore, the electro-osmotic mobility is usually determined by the streaming current or streaming potential measurement.

When a narrow rectangular or circular capillary is considered, the solution volume flux ($Q$) and electric current ($I$) through the capillary can be expressed using the applied pressure difference ($\Delta P$) and applied voltage ($\Delta\Psi$) as follows:
\begin{eqnarray}
Q &=& K \Delta P + \mu \frac{S}{L} \Delta \Psi,\\
I &=& \mu \frac{S}{L} \Delta P + G \Delta \Psi, 
\end{eqnarray}  
where $K$ is the permeation (Darcy) coefficient and $G$ is the conductance.
The electro-osmotic coefficient is expressed as the product of the electro-osmotic mobility ($\mu$) and $S/L$, where $S$ is the cross-section and $L$ is the length of the capillary. 
The coefficient of the pressure-induced current (streaming current) is the same as the electro-osmotic coefficient because of the Onsager reciprocal relation.
If the electric circuit is closed, the current under the applied pressure difference is the streaming current, whereas if it is open ($I=0$), the electrostatic potential under the applied pressure difference is the streaming potential, which is expressed as
\begin{equation}
\Delta\Psi = -\frac{\mu}{G}\frac{S}{L}\Delta P.
\end{equation}
In general, the electro-osmotic mobilities of hydrophobic PTFE/water interfaces are derived from streaming current and streaming potential measurements \cite{Werner1998, Barii2021,Gulgonul2019}.
However, the electro-osmotic mobilities of liquid/water and air/water interfaces cannot be determined by electro-osmosis or streaming current (potential) measurements because the fluid interface is mobile and such experiments are difficult to define.
To the best of our knowledge, electro-osmosis of air/water interfaces has been investigated only in a few studies \cite{Usui1987,HusseinSheik2017,Blanc2018}, and no study has been conducted on the electro-osmosis of liquid/water interfaces thus far. 
Therefore, the zeta potentials of the gas/water and liquid/water interfaces are usually calculated from the electrophoretic mobility of liquid droplets or bubbles. 

Zeta potential of the interface can also be calculated from the electrophoretic mobility measurements of the particles. 
When the particle radius is much larger than the Debye length ($\kappa R\gg 1$), the electrophoretic mobility can be expressed as 
\begin{equation}
\mu = \frac{\varepsilon\varepsilon_0}{\eta}\zeta 
\label{eq:21}
\end{equation}
where $\zeta$ is the arbitrary strength.
This equation is called the Helmholtz-Smoluchowski formula and is useful for obtaining zeta potentials. 

\begin{figure}
\includegraphics[width=8cm]{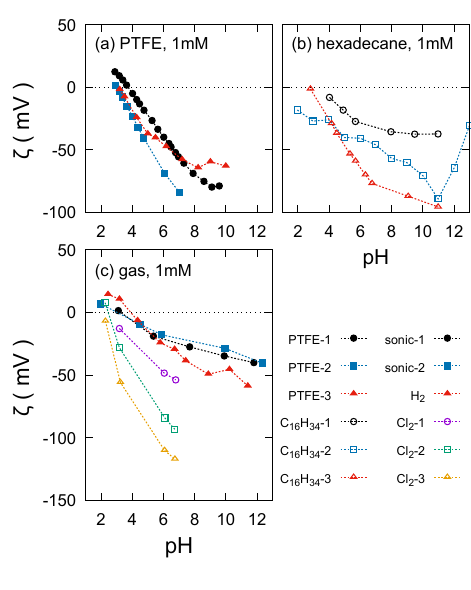}
\caption{Zeta potential of various hydrophobic interfaces as a function of $\mathrm{pH}$ in 1 mM electrolyte solutions.
(a) Zeta potentials of PTFE/water interfaces: 
PTFE-1: flat surface of plasma-deposited PTFE in KCl solution \cite{Werner1998}, 
PTFE-2: flat surface of PTFE in NaCl solution \cite{Barii2021}, 
PTFE-3: flat surface of PTFE in NaCl solution \cite{Gulgonul2019}.
(b) Zeta potential of hexadecane/water interfaces:
$\mathrm{C_{16}H_{34}}$-1: hexadecane droplets (99.8\% purity, $R = 100\,$nm) in NaCl solution \cite{Roger2012}, 
$\mathrm{C_{16}H_{34}}$-2: hexadecane droplets ($R = 125\,$nm) in NaCl solution \cite{Yang2017}, 
$\mathrm{C_{16}H_{34}}$-3: hexadecane droplets ($R > 1\,\mu$m) in NaCl solution \cite{Michalek1996}, 
(c) Zeta potential of gas/water interfaces: 
sonic-1:  bubbles ($R$ is unknown) generated by ultrasonic irradiation in KCl solution \cite{Elmahdy2008}, 
sonic-2: bubbles ($R=400\,$nm) generated by ultrasonic irradiation in NaCl solution \cite{Cho2005}, 
$\mathrm{H_2}$: bubbles ($R>5\,\mu$m) generated by electrolysis in NaCl solution \cite{Yang2001}, 
$\mathrm{Cl_2}$-1 to 3: bubbles with radius $R = 30\,\mu$m (open circles), $100\,\mu$m (open squares), and $150\,\mu$m (open triangles) generated by electrolysis in KCl solution \cite{Smith1985}.
}
\label{fig:8}
\end{figure}

Fig.~\ref{fig:8} shows the zeta potentials of hydrophobic solid \cite{Werner1998,Barii2021,Gulgonul2019}, liquid \cite{Roger2012,Yang2017,Michalek1996}, and gaseous \cite{Elmahdy2008,Cho2005,Yang2001} interfaces as a function of pH in 1 mM NaCl or KCl solutions.
All the plots exhibit an isoelectric point at $\mathrm{pH}=2$ to $4$, and the pH dependence for all samples is similar. 
In Fig.~\ref{fig:8}a, all solid surfaces were flat PTFE, and the zeta potentials were calculated from the streaming current (or potential). 
The experimental data fit well into one curve, even though the experimental systems and materials are different. 
In contrast, the zeta potentials of hexadecane/water and gas/water interfaces are scattered, as shown in Fig.~\ref{fig:8}b and c even though the salt concentrations are the same.
This indicates that eq.~\ref{eq:21} is inaccurate for liquid droplets and gas bubbles in water. 
For example, the three lines with open symbols in Fig.~\ref{fig:8}c denote the same experimental system \cite{Smith1985}, but the diameters of the $\mathrm{Cl_2}$ bubbles are different ($d=30\,\mu$m, $100\,\mu$m, and $150\,\mu$m). 
Experiments suggest that the zeta potential of the gas bubble depends linearly on the radius of the bubble, as $\mu = \mu_0 + \mu_1 \cdot \kappa R$, where $\mu_0$ and $\mu_1$ are arbitrary constants \cite{Smith1985}, which cannot be explained by applying a more rigorous theory of solid particle electrophoresis \cite{Henry1931,OBrien1978}.
This dependence of the zeta potential on the electrophoretic mobility is similar to that proposed for a mercury droplet \cite{Ohshima1984}, and is linked to the Maxwell stress at the boundary \cite{Ohshima1984, Uematsu2022}.
To date, the linear dependence of the zeta potential of gas bubbles on their radius has not been fully explained.

\section{Conclusion}

In this chapter, experimental data of contact angles, surface/interfacial tensions, and zeta potentials have been used to investigate whether all hydrophobic interfaces have the same universal properties.

Contact angles of solid surfaces can be measured and this concept can be extended to liquid/water interfaces using the surface/interfacial tensions of liquid/air and liquid/water interfaces.
However, this concept cannot be extended to a gaseous hydrophobic interface, because interfacial tension at the gas/air interface cannot be measured. 

Ion adsorption at a hydrophobic interface can be determined using the surface/interfacial tension difference caused by the addition of salts, 
as reported in the literature; this data will help in determining the affinity strengths of cations, anions, and the water autoionization ions, $\mathrm{H^+}$ and $\mathrm{OH^-}$. 
The pristine air/water interface is confirmed to be positively charged based on a careful analysis of the experimental data of the surface tension differences.
Compared to those of the air/water interface, reports of the interfacial tension differences of the liquid/water interface caused by the addition of electrolytes, acids, and bases are limited in the literature. 
We expect that the affinity strengths of ions for hydrocarbon/water interfaces are quantitatively similar to those for air/water interfaces; however, more experimental data are necessary to validate this inference.  
The interfacial tension differences of the solid/water interfaces caused by the addition of salts can be calculated from the experimental contact angle data. 
However, reports of the contact angles of electrolyte solutions on hydrophobic solid surfaces are few in the literature, and it remains unclear whether the affinity strength of $\mathrm{H^+}$ and $\mathrm{OH^-}$ for the hydrophobic solid/water interface is quantitatively similar to that for the air/water interface. 

The zeta potential of a hydrophobic interface is usually measured by streaming current (or potential) or electrophoresis. 
All experimental data exhibit similar behavior, i.e., the isoelectric point is located at $\mathrm{pH}=2$ to 4, and the zeta potential is negative at neutral and basic pH.
When the zeta potentials of the solid/water, liquid/water, and gas/water interfaces were compared, the zeta potential of PTFE/water interfaces fit well into one curve, whereas those of the liquid/water and gas/water interfaces were scattered; 
this is because of the incorrect application of the Helmholtz-Smoluchowski formula to the electrophoresis of liquid droplets and bubbles. 

Important aspects of hydrophobic interfaces such as friction properties \cite{Huang2008}, interactions between two interfaces \cite{Israelachvili1982}, the effect of dissolved gases \cite{Ishida2000}, and spectroscopic methods for detecting ion adsorption at hydrophobic interfaces \cite{Shen2006} have not been discussed in this chapter.
We believe that this chapter provides a fundamental understanding of the electric properties of hydrophobic interfaces. 

\section*{Acknowledgements}

This work was supported by JST, PRESTO Grant Number JPMJPR21O2.
The author is thankful for JSPS KAKENHI Grants 18KK0151, 20K14430, 22K03546, and a grant from the Kurita Water and Environment Foundation (21E006).
The author also thanks H. Shiba for critical reading of the manuscript. 

\bibliography{ref}

\end{document}